\documentclass[aps,amsmath,amssymb,amsfonts,twocolumn,prl,showpacs]{revtex4}
\usepackage{graphicx}
\usepackage{bm}
\usepackage{color}

\newcommand{\mbf}{\mathbf}
\newcommand{\mrm}{\mathrm}

\begin{document}
\author{Zbigniew Idziaszek}
\affiliation{Faculty of Physics, University of Warsaw, 00-681 Warsaw, Poland}
\author{Paul S. Julienne}
\affiliation{Joint Quantum Institute, NIST and
the University of Maryland, Gaithersburg, Maryland 20899-8423, USA}
\title{Universal rate constants for reactive collisions of ultracold molecules}

\begin{abstract}
A simple quantum defect model gives analytic expressions for the complex scattering length and threshold collision rates of ultracold molecules.  If the probability of reaction in the short-range part of the collision is high, the model gives universal rate constants  for $s$- and $p$-wave collisions that are independent of short-range dynamics.  This model explains the magnitudes of the recently measured rate constants for collisions of two ultracold $^{40}$K$^{87}$Rb molecules, or an ultracold $^{40}$K atom with the $^{40}$K$^{87}$Rb molecule [S. Ospelkaus {\it et al.}, Science {\bf 327}, 853 (2010)].
\end{abstract}

\pacs{03.65.Nk, 34.10.+x, 34.50.Cx, 34.50.Lf}

\maketitle

Research utilizing ultracold atoms has become established as a major forefront multidisciplinary area  involving diverse areas such as atomic and molecular physics, quantum optics, condensed matter physics, and quantum information.   Extending such research with ultracold molecules will open up a number of new opportunities~\cite{Doyle2004,Carr2009}.  Development of sources of ultracold molecules is progressing rapidly~\cite{Faraday2009}.  It is essential to understand the elastic, inelastic, and reactive collisions of such molecules  in order to control and utilize them effectively.  A recent experiment reported the first evidence of ultracold ``chemistry'' and measured the reactive collision rates of fermionic $^{40}$K$^{87}$Rb molecules between 250 and 900 nK~\cite{Ospelkaus2009}. Since molecules are much more complex than atoms, it is necessary to develop models of threshold molecular collisions that are adequate to account for their complexity yet simple enough for understanding experimental data.

Here we propose a simple, yet general, model for threshold molecular collision rates based on  the separation of the effects of the long- and short-range parts of the intermolecular potential for a given scattering channel~\cite{JulienneFaraday2009}.  Our fully quantum mechanical model uses the analytic framework of generalized multichannel quantum defect theory (MQDT)\cite{SeatonRPP1983,GreenePRA1982} based on the specific version pioneered by Mies~\cite{MiesJChPh1984,MiesJChPh1984b,Julienne1989}.  The reactive or inelastic part of the short-range collision is characterized by a dimensionless ``quantum defect'' parameter $ 0\le y \le 1$, which is related to the probability of irreversible loss of incoming scattering flux from the entrance channel due to dynamics at short range.  The role of the long-range potential is to determine how much of the entrance channel wave with very large de Broglie wave length is transmitted to short range to experience such loss dynamics.  Our MQDT model generalizes the model of  Ref.~\cite{OrzelPRA1999}, which has been successfully used to understand the magnitude of inelastic collision rates involving ultracold atoms.

We find universal rate constants for $s$- and $p$-wave collisions depending only on the long-range potential when $y \to 1$, which corresponds to unit probability of reaction at short range so that no flux is reflected at short range back into the entrance channel.   In contrast, if $y  \ll 1$, the rate constants depend strongly on the scattering length of the entrance channel and are not universal.

We consider collisions of two particles interacting via the van der Waals (vdW) potential at long-range. These can be $S$-state atoms, or molecules in the rovibrational and electronic ground state. The scattering channels $|\alpha\rangle = |a_1 a_2\rangle |\ell m \rangle$ are defined in terms of their
respective internal states $a_1$ and $a_2$, and the partial wave quantum numbers $\ell m$.
In general, the elastic rate constant $\mathcal{K}^\mrm{el}$ and the rate constant  $\mathcal{K}^\mrm{ls}$ for total inelastic or reactive scattering are given by the diagonal elements $S_{\alpha\alpha}$ of the $S$-matrix for channel $\alpha$.   Expressing $S_{\alpha\alpha} = e^{2 i \eta_{\alpha}}$ in terms of
an energy-dependent complex scattering length $\tilde{a}_{\alpha}(k) = \tilde{\alpha}_{\alpha}(k) - i \tilde{\beta}_{\alpha}(k)$ \cite{HutsonNJP2007}, defined analogously  to the $s$-wave energy-dependent scattering length~\cite{BlumePRA2002,BoldaPRA2002},
\begin{equation}
\label{atilde}
\tilde{a}_{\alpha}(k_\alpha) = - \frac{\tan \eta_{\alpha}(k_\alpha)}{k_\alpha} = \frac{1}{ik_\alpha} \frac{1 -S_{\alpha\alpha}(k_\alpha)}{1+S_{\alpha\alpha}(k_\alpha)}\,,
\end{equation}
gives the rate constant contributions from channel $\alpha$,
\begin{align}
\mathcal{K}^\mrm{el}_\alpha(E) & = g_\alpha \frac{\pi \hbar}{\mu k_\alpha} \left| 1 - S_{\alpha\alpha}(E) \right|^2 = 2 g_\alpha\frac{h k_\alpha}{\mu} |\tilde{a}_{\alpha}(k_\alpha)|^2 f_{\alpha}(k_\alpha) \,,   \label{Kel} \\
\mathcal{K}^\mrm{ls}_\alpha(E) & = g_\alpha\frac{\pi \hbar}{\mu k_\alpha} \left(1- |S_{\alpha\alpha}(E)|^2\right) =
2g_\alpha\frac{h}{\mu} \tilde{\beta}_{\alpha}(k_\alpha) f_{\alpha}(k_\alpha) \,, \label{Kloss}
\end{align}
where $k_\alpha^2 =2 \mu (E-E_\alpha)/\hbar^2$ with $E$ denoting the total energy, $\mu$ the reduced mass, and $E_\alpha$ the threshold energy of the channel $\alpha$.   The factor $g_\alpha=1$ except that $g_\alpha=2$ when both particles are identical species in identical internal states, $a_1=a_2$; $\ell$ is restricted to being even (odd) in the case of identical bosons (fermions).  The function
\begin{equation}
\label{fellm}
f_{\alpha}(k_\alpha) = \frac{1}{1+k_\alpha^2|\tilde{a}_{\alpha}(k_\alpha)|^2+2k_\alpha\tilde{\beta}_{\alpha}(k_\alpha)},
\end{equation}
has the property that $0< f_{\alpha}(k_\alpha)\le 1$ and $ f_{\alpha}(k_\alpha) \to 1$ as $k_\alpha \to 0$.  The latter is true when both conditions $k_\alpha|\tilde{a}_{\alpha}(k_\alpha)| \ll 1$ and $k_\alpha\beta_{\alpha}(k_\alpha) \ll 1$ are met. We stress that Eqs.~\eqref{Kel}-\eqref{Kloss} are {\it exact} \cite{HutsonNJP2007}, and they apply either to coupled channel (unitary) and complex potential (non-unitary) models.

We introduce a single channel model with a complex potential to represent the long range potential and short range loss dynamics for channel $\alpha$, where for simplicity of notation we drop the implied channel index $\alpha$:
\begin{equation}
U_\ell(r) = V(r) + \frac{\hbar^2 \ell(\ell+1)}{2 \mu r^2} - i \frac{\gamma(r)}{2} \,,
\end{equation}
and $V(r)$ has the vdW form at long-range: $V(r) = - C_6/r^6$ for $r \gtrsim R_0$, where $R_0$ denotes the range of short-range forces (e.g. the exchange interaction).  The vdW potential is characterized by the length $R_6 = \frac12 \left(2 \mu C_6/\hbar^2\right)^{1/4}$, or the closely related length $\bar{a} = 4 \pi R_6/\Gamma(\frac14)^2 $~\cite{GribakinPRA1993}.  The imaginary part $\gamma(r)$, which is assumed to vanish beyond  $R_0$, simulates all short-range coupling at $r < R_0$ to
exoergic non-threshold exit channels that result in loss from the entrance channel. We assume that the kinetic energy release in loss channels is much larger than any exit barriers. The short range $V(r)$ for $r < R_0$ should be viewed as a pseudopotential which determines the phase of the wave function for $r >R_0$.  This phase determines the ``background'' s-wave scattering length $a$ for the potential $V(r)$, which can in general be different for different $\ell m$ in the case of general anisotropic potentials.

Decomposing the wave function into real and imaginary parts, $\Psi(r) = F(r) + i G(r)$, gives
the two channel Schr\"odinger equation
\begin{equation}
\label{RadialSchr}
\frac{\partial^2 \mbf{\Phi}}{\partial r^2} + \frac{2 \mu}{\hbar^2}\left(E - \mbf{W}(r) \right) \mbf{\Phi}(r) = 0,
\end{equation}
for the wave function $\mbf{\Phi}(r) = \{F(r),G(r)\}$ with a non-Hermitian interaction matrix
\begin{equation}
\mbf{W}(r) = \left( \begin{array}{cc}
V(r) + \frac{\hbar^2 \ell(\ell+1)}{2 \mu r^2} & \frac{\gamma(r)}{2} \\
-\frac{\gamma(r)}{2} & V(r) + \frac{\hbar^2 \ell(\ell+1)}{2 \mu r^2}
\end{array} \right).
\end{equation}
We solve the two-channel problem in the framework of MQDT \cite{SeatonRPP1983,GreenePRA1982,MiesJChPh1984}, which separates the effects of the long- and short-range parts of the potential. We adopt the notation of Mies \cite{MiesJChPh1984}, and introduce a quantum defect matrix $\mbf{Y}(E) = \{\{0,-y(E)\},\{y(E),0)\}\}$.  The antisymmetric form of $\mbf{Y}$ results from the non-Hermitian potential $\mbf{W}(r)$.

The wave function for $r \gtrsim R_0$ is
\begin{equation}
\mbf{\Phi}(r) = \left[ \hat{f}(r) \mbf{I}\ +  \hat{g}(r) \mbf{Y}(E) \right] \mbf{A} \quad .
\end{equation}
Here, $\mbf{A}$ is some constant vector, $\mbf{I}$ is the identity matrix, and $\hat{f}(r)$,  $\hat{g}(r)$ are solutions of the Schr\"odinger equation in the real part of the potential $V_\ell(r)$ that have ``local'' WKB-like normalization at short distances \cite{MiesJChPh1984, Julienne2009b},
\begin{align}
\label{fghat}
\left.
\begin{array}{lll}
\hat{f}(r,E) & \cong & k(r)^{-1/2} \sin \beta (r),\phantom{\Big(}\\
\hat{g}(r,E) & \cong & k(r)^{-1/2} \cos \beta (r),\phantom{\Big(}\\
\end{array}
\right\}\quad r \gtrsim R_0,
\end{align}
where $k(r) = \sqrt{2 \mu \left(E - \mrm{Re}[U_\ell(r)]\right)}/\hbar$ is the local wave vector, and $\beta(r) = \int^r\!\mrm{d}x\,k(x)$ is the WKB phase.  Typically, $R_0 \ll R_6$, and for $s$- or $p$-wave ultracold collision one can assume that $y(E) = y$ is independent of $E$.  The two parameters $a$ and $y$ completely represent the effects of the complex short-range dynamics on the wave function in channel $\alpha$ for $r \gtrsim R_0$.

Some intuition about $y$ is gained from writing $\Psi(r)$ at small distances, $R_0 < r \ll R_6$, using the WKB-like form in Eq.~\eqref{fghat} for $\hat{f}(r)$ and $\hat{g}(r)$:
\begin{equation}
\Psi(r) \sim \frac{\exp\left[-i \int^r\!k(x)\mrm{d}x\right]}{\sqrt{k(r)}} - \left(\frac{1-y}{1+y}\right) \frac{\exp\left[i \int^r\!k(x)\mrm{d}x\right]}{\sqrt{k(r)}}.
\end{equation}
The first term represents the flux of incident particles, whereas the second term gives the flux
reflected from the short-range potential. Hence, for $y=1$ there is no outgoing flux, whereas for $y$ not unity there is some back reflection. Finally, for $y=0$ (no losses), the incident and reflected fluxes are equal, and we recover the standard scattering wave function $\Psi(r) \sim k^{-1/2}(r) \sin \left[\int^r\!k(x)\mrm{d}x\right] = \hat{f}(r)$. In the following we will assume $0 \le y \le 1$~\cite{footnote1}.

Applying the standard MQDT formulas relating the quantum-defect matrix $\mbf{Y}$ to the scattering
S-matrix \cite{MiesJChPh1984}, we calculate $\tilde{a}(E)$ assuming $k R_6 \ll 1$
\begin{align}
\label{acomp}
\tilde{a}(E) & \approx -\frac{1}{k} \left( \tan \xi(E) - \frac{y C^{-2}(E)}{i+y \tan \lambda(E)} \right),
\end{align}
Here, $C(E)$ and $\tan \lambda(E)$ are the MQDT functions \cite{MiesJChPh1984} that connect the solutions with short- and long-range normalization: $f(r) = C^{-1}(E) \hat{f}(r)$, $g(r) = C(E)[\hat{g}(r)+\tan \lambda(E) \hat{f}_i(r)]$, where the scattering solutions in the asymptotic region are defined as follows:
\begin{align}
\label{fg}
\left.
\begin{array}{ll}
f(r,E) \cong \sin \left(kr-\ell\pi/2+\xi\right)/\sqrt{k},\\
g(r,E) \cong \cos \left(kr-\ell\pi/2+\xi\right)/\sqrt{k},
\end{array}
\right\}\quad r \rightarrow \infty,
\end{align}
where $\xi(k)$ is the phase shift.  The functions $C(E)$, $\tan \lambda(E)$ and $\xi(E)$ are found from the analytic vdW theory~\cite{GaoPRA1998}.  The small-$k$ behavior for an $s$ wave is
\begin{align}
\label{MQDTFunthr1}
C^{-2}(E,\ell =0) & \stackrel{k \rightarrow 0}{\longrightarrow} k \bar{a}
\left(1+ (s-1)^2\right), \\
\label{MQDTFunthr2}
\tan \lambda(E,\ell =0) & \stackrel{k \rightarrow 0}{\longrightarrow} 1-s,
\end{align}
and $\tan \xi(E,\ell = 0) \stackrel{k \rightarrow 0}{\longrightarrow} - k a$, where $s=a/\bar{a}$ is the dimensionless scattering length. For a $p$ wave:
\begin{align}
\label{CePThr}
C^{-2}(E,\ell=1) & \stackrel{k \rightarrow 0}{\longrightarrow}
2 k \bar{a}_1 (k \bar{a})^2 \frac{1+(s-1)^2}{(s-2)^2},\\
\label{TanPThr0}
\tan \lambda(E,\ell=1) & \stackrel{k \rightarrow 0}{\longrightarrow} s/(s-2),\\ \label{TanPThr1}
\tan \xi (E,\ell=1) & \stackrel{k \rightarrow 0}{\longrightarrow} 2 k \bar{a}_1 (k \bar{a})^2 \frac{s-1}{s-2},
\end{align}
where $\bar{a}_1 = \bar{a} \Gamma(\frac14)^6/(144 \pi^2 \Gamma(\frac34)^2)  \approx 1.064 \bar{a}$.

For small $y$ Eqs.~\eqref{Kloss} and \eqref{acomp} predict that the loss rate scales as $\tilde{\beta}(E) \approx y C^{-2}(E)$. This can be interpreted~\cite{Julienne1989} as an energy-insensitive probability $y$ of the short-range reaction multiplied by the function $C^{-2}(E)$, which provides the proper scaling of the wave function amplitude between the short-range and asymptotic zones.

A key result of this paper is Eq.~\eqref{acomp} for $\tilde{a}(E)$ for arbitrary $y$.  Equations \eqref{MQDTFunthr1}-\eqref{TanPThr1} yield the small-$k$ analytic limits of  Eq.~\eqref{acomp}, valid when $k|\tilde{a}| \ll 1$ and $k R_6 \ll 1$:
\begin{align}
\label{al0}
\tilde{a}_{\ell=0}(k) & = a + \bar{a} y \frac{1+(1-s)^2}{i+y(1-s)}, \\
\label{al1}
\tilde{a}_{\ell=1}(k) & = - 2 \bar{a}_1 (k \bar{a})^2 \frac{y+i (s-1)}{ys +i(s-2)}\,.
\end{align}
These expressions parameterize the elastic and inelastic collision rates in terms of the two dimensionless parameters $s$ and $y$,  namely, an entrance channel phase and a short-range interchannel coupling strength, plus the van der Waals parameter $\bar{a}$.  In the special case of unit probability of short range loss from the entrance channel, $y \to 1$, we find the universal result that $\tilde{\alpha}$ is {\it independent} of $s$ and $y$, and depends on $\bar{a}$ only.  Specifically, $\alpha_{\ell=0} = \beta_{\ell=0} = \bar{a}$  for an $s$-wave and $\alpha_{\ell=1} = - \bar{a}_1 (k \bar{a})^2$ and $\beta_{\ell=1} \rightarrow \bar{a}_1 (k \bar{a})^2$ for a $p$-wave.  This gives the universal rate constants determined solely by the quantum transmission of the long-range potential:
\begin{align}
\label{Kl0}
&\mathcal{K}^\mrm{el}_{\ell=0}  =  4 g \frac{h}{\mu} k \bar{a}^2& &\mathcal{K}^\mrm{ls}_{\ell=0}  =  2 g \frac{h}{\mu} \bar{a}\,\,\,\\
\label{Kl1}
&\mathcal{K}^\mrm{el}_{\ell=1}  =  4 \sigma g \frac{h}{\mu} k \bar{a}_1^2 (k\bar{a})^{4} & &\mathcal{K}^\mrm{ls}_{\ell=1}  = 2 \sigma g \frac{h}{\mu} \bar{a}_1(k\bar{a})^2
\end{align}
valid for $k R_6 \ll 1$. The factor $\sigma = 3$ for $\ell=1$ for a rotationless molecule in the universal limit, since all three $m$ components have the same $C_6$ and contribute equally.  In general, a sum of the different contributions from each $m$ would need to be taken.

We have verified Eqs.~\eqref{Kl0} and~\eqref{Kl1} by numerical calculations with a complex potential.  These equations also reproduce the numerical model of Orzel {\it et al}~\cite{OrzelPRA1999} for $s$- and $p$-waves, for which $y=1$ is an excellent approximation for ionizing collisions of Xe metastable atoms.  Ref.~\cite{OrzelPRA1999} found the ionization rate constant for $s$-wave collisions to be independent of isotopic mass, that is, independent of $s$, which is expected to be quite different for different isotopes~\cite{Julienne2009b}.  The value of $\mathcal{K}^\mrm{ls}_{\ell=0}$ predicted by Eq.~\eqref{Kl0} agrees with the measured value of Ref.~\cite{OrzelPRA1999} within experimental uncertainty.    Hudson {\it et al.}~\cite{Hudson2008} also found that a numerical implementation of the model of Ref.~\cite{OrzelPRA1999} explained their vibrational relaxation data for RbCs molecules, although the error bars were quite large.

\begin{figure}[tb]
	 \includegraphics[width=8.6cm,clip]{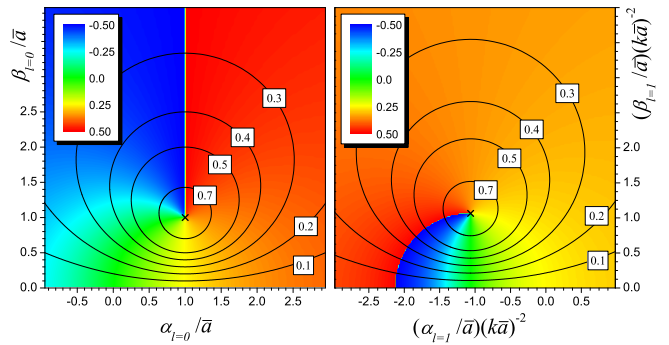}
	 \caption{Real and imaginary parts of $\tilde{a}_{\ell=0}/\bar{a}$ (left panel) and $(\tilde{a}_{\ell=1}/\bar{a})(k\bar{a})^{-2}$ (right panel) as $k \to 0$ for different values of the loss parameter $y$ (set of circles) as the phase parameter $s=a/\bar{a}$, color coded according to the value of $\arctan(s)/\pi$, varies over its full range of $-\infty < a < +\infty$.  The blue-red boundary occurs where $U_\ell(r)$ has a bound state at $E=0$.
	 \label{fig:circles}
	 }
\end{figure}

In general, $y$ is not unity, and there can be reflected flux from short-range back into the entrance channel, so that $\tilde{a}$ is sensitive to both $s$ and $y$.  Figure~\ref{fig:circles} shows the complex scattering length $\tilde{a}$ plotted in the $\{\alpha,\beta\}$ plane as $y$ and $s$ vary for $s$- and $p$-waves.  The Figure shows $\tilde{a}$ lies on a circle for a fixed values of $y$. For $s$-waves these circles are centered at $\{\bar{a},\bar{a}(\frac{y}{2}+\frac{1}{2y})\}$, and have radius $\bar{a}(1-y^2)/(2 y)$, while for $p$-waves they are centered at $\{-\bar{a}_1,\bar{a}_1(\frac{y}{2}+\frac{1}{2y})\}$ and have radius $\bar{a}_1(1-y^2)/(2 y)$. The points at the top (bottom) of the circles correspond to the maximal (minimal) loss: $\beta_{\ell =0} = \bar{a}/y$ and $\beta_{\ell=1} = \bar{a}_1/(y k^2 \bar{a}^2)$ ($\beta_{\ell=0} = y \bar{a}$ and
$\beta_{\ell=1} = y \bar{a}_1/(k^2 \bar{a}^2)$), and they are realized at $s=\infty$ and $s=2$~\cite{footnote2}  ($s=1$ and $s=0$), for the $s$- and $p$-wave, respectively. Hence, counterintuitively, the largest loss rates ($\beta \gg \bar{a}$) occur only for small values of the reaction probability $y$, but only when $s>1$ is large enough that there is a near-threshold bound state that allows the $C(E)^{-2}$ function to build up a large amplitude of the short-range wave function.

\begin{figure}[tb]
	 \includegraphics[width=7cm,clip]{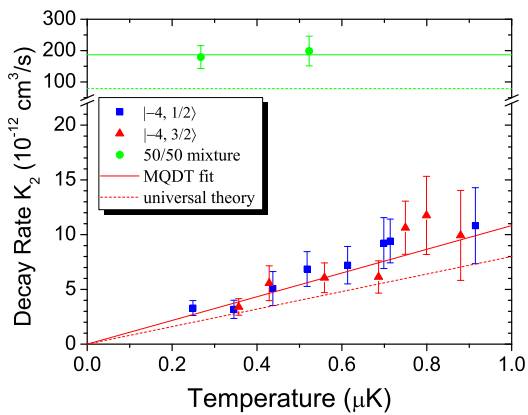}
	 \caption{Loss rate constant ${\cal K}^\mrm{ls}$ of KRb molecules versus temperature. Experimental data \cite{Ospelkaus2009} for a single component gas of molecules in spin states $|F=-4,M_F =\frac12\rangle$ (blue squares), $|F=-4,M_F =\frac12\rangle$ (red triangles), and for 50/50 mixture of these two spin states (green dots), are compared with prediction of MQDT in the universal regime ($y=1$, dashed lines), and in non universal regime ($y<1$, solid lines). In the latter case the values of $y$, and $s$ parameters for $\ell=0$ and $\ell=1$ are determined by fitting to the experimental data.
	 \label{fig:JILA}
	 }
\end{figure}

Our theory explains recent experimental data on the reaction rate coefficients of a gas of ultracold $^{40}$K$^{87}$Rb molecules in their vibrational and rotational ground state~\cite{Ospelkaus2009}.   These molecules are fermions with 36 distinct states of nuclear spin.  The experiment could prepare the molecules in either the same or different spin levels, thereby allowing measurement of both $s$- and $p$-wave reactive collision rates.  Fig.~\eqref{fig:JILA} compares the experimental data with two different predictions of our MQDT model.   The dashed line shows the predictions of the universal model at low $k$, where $\mathcal{K}^\mrm{ls}_{\ell=0}(T)=4(h/\mu)\bar{a}$ is independent of temperature $T$ and
\begin{align}
  \mathcal{K}^\mrm{ls}_{\ell=1}(T) = \frac{\Gamma(1/4)^6}{\Gamma(3/4)^2} \bar{a}^3 \frac{k_B T}{h} = 1513  \bar{a}^3 \frac{k_B T}{h} \,.
\end{align}
varies linearly with $T$.  With $\bar{a}=118(3)$ a$_0$ from two {\it ab initio} calculations of
$C_6$~\cite{Kitochigova2009,Moszynski2009}, assuming a 5 $\%$ uncertainty, $ \mathcal{K}^\mrm{ls}_{\ell=1}(T)/T=0.8(1)\times 10^{-5}$ cm$^3$s$^{-1}$K$^{-1}$, compared to the measured value of
$1.1(3)\times 10^{-5}$ cm$^3$s$^{-1}$K$^{-1}$ for the lowest spin state~\cite{Ospelkaus2009}.

While the universal $p$-wave $ \mathcal{K}^\mrm{ls}_{\ell=1}$ agrees with the data within mutual uncertainties, the $s$ wave $\mathcal{K}^\mrm{ls}_{\ell=0}=0.8\times 10^{-10}$ cm$^3/$s is a factor of 2 smaller than the measured $1.9(4)\times 10^{-10}$ cm$^3/$s.   This departure from universality may yield information about the short range dynamics. We have fitted the formula \eqref{Kloss} with $\tilde{\alpha}$ and $\tilde{\beta}$ from Eqs.~\eqref{al0} and \eqref{al1} to both the $s$- and $p$-wave experimental data to estimate values of the $y$ and $s$ parameters, assuming two distinct $s$ parameters for $\ell=0$ and $\ell=1$.   Since the equations for the complex scattering length are quadratic in $s$, this yields two sets of solutions: (i) $y = 0.397$, $s_{\ell=0} = 10.3$, $s_{\ell=1} = 3.43$; (ii) $y = 0.420$, $s_{\ell=0} = -48$, $s_{\ell=1} = 3.43$,   The solid lines in Fig.~\eqref{fig:JILA} show that agreement can be obtained with the data in this way, implying that a significant fraction of the incoming flux may be reflected back into the entrance channel.

Our model predictions can be tested by other experiments.  For example, if $y$ is not unity, unlike the case studied by Ref.~\cite{OrzelPRA1999}, there would be some dependence on isotopic mass, since $s$ will vary with mass~\cite{GribakinPRA1993,Julienne2009b}.  Thus, other isotopic species would depart from the universal rate constants differently than the $^{40}$K$^{87}$Rb isotope.  In addition, experiments with an electric field turned on to induce a molecular dipole should have short range reactions governed by the same $s$ and $y$ parameters, if the asymptotic solutions of the dipole problem are matched to the solutions in the vdW zone $R_0 < r < R_6$.  Thus, collision rate measurements in an electric field could test universality and the role of short range dynamics.

Equation~\eqref{Kl1} predicts a universal rate constant of $1.1(1)\times 10^{-10}$ cm$^3/$s for reactive $s$-wave collision of $^{40}$K atoms with $^{40}$K$^{87}$Rb molecules, using the $C_6$ constant from Kotochigova~\cite{Kitochigova2009}.  This is quite close to the measured value, $1.7(3)\times 10^{-10}$ cm$^3/$s, of Ospelkaus {\it et al}~\cite{Ospelkaus2009}.

Our MQDT parameterization provides a simple framework for characterizing and understanding threshold molecular collisions.  The MQDT method could readily be generalized to include the effects of molecular resonance states, threshold exit channels, finite electric fields, or reduced dimensional geometry.    Collisional resonances should be suppressed when $y \to 1$ but should be prominent if $y \ll 1$.  A MQDT framework may also offer a way to incorporate quantum dynamics into simpler generalized quantum threshold Langevin models of cold molecular collisions in electric fields~\cite{Quemener2009}.

We acknowledge support from an AFOSR MURI on Ulracold Molecules and a Polish Government Research Grant for 2007-2009.  We thank Svetlana Kotochigova, Robert Moszy\'nski, John Bohn, and Goulven Qu\'em\'ener for discussions on theory, and Jun Ye and Deborah Jin for providing the JILA experimental data.

\bibliography{molMQDT}

\begin{thebibliography}{23}
\bibitem[{foo({\natexlab{a}})}]{footnote1}
\bibinfo{note}{The fact that the reflected flux must be no greater than the
  incident flux requires $y>0$. For $y>1$ the prefactor $-1<(1-y)/(1+y)<0$ can
  be mapped to the case of $0 \leq y \leq 1$ with a different WKB phase. This
  can be formally shown using the transformation: $y \to 1/y$ and $a \to
  \bar{a}(2\bar{a}-a)/(\bar{a}-a)$.}
\end{thebibliography}
\end{document}